\documentclass{moriond}

\usepackage{hyperref}
\usepackage{amssymb}
\def\Journal#1#2#3#4{{#1} {\bf #2}, #3 (#4)}

\def\PR{{\em Phys. Rev.}}
\def\EPJST{{\em EPJ S. T.}}
\def\EPJC{{\em EPJ C}}

\def\be{\begin{equation}}
\def\ee{\end{equation}}
\def\bea{\begin{eqnarray}}
\def\eea{\end{eqnarray}}

\newcommand{\lambdabar}{{\mkern0.75mu\mathchar '26\mkern -9.75mu\lambda}}

\newcommand{\Photo}{\includegraphics[height=35mm]{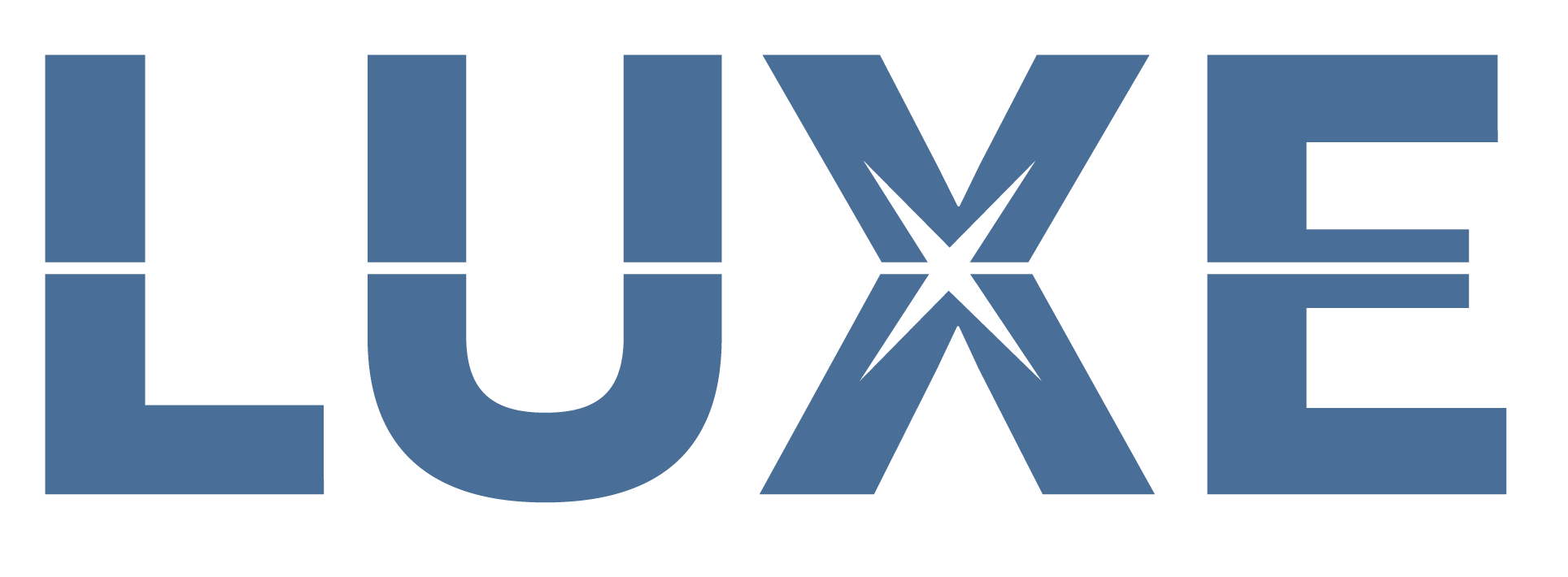}}

\begin{document}
\vspace*{4cm}
\title{LUXE: A new experiment to study non-perturbative QED in $e^-$-laser and $\gamma$-laser collisions.}

\author{ Ruth Jacobs, on behalf of the LUXE collaboration }

\address{Deutsches Elektronen-Synchrotron DESY, Notkestr. 85, 22607 Hamburg, Germany}

\maketitle\abstracts{
The LUXE experiment (Laser Und XFEL Experiment) is a new experiment in planning at DESY Hamburg using the electron beam of the European XFEL. At LUXE, the aim is to study collisions between a high-intensity optical laser and up to $16.5\,$GeV electrons from the Eu.XFEL electron beam, or, alternatively, high-energy secondary photons. The physics objectives of LUXE are to measure processes of Quantum Electrodynamics (QED) at the strong-field frontier, where QED is non-perturbative. This manifests itself in the creation of physical electron-positron pairs from the QED vacuum. LUXE intends to measure the positron production rate in a new physics regime at an unprecedented laser intensity. Additionally, the high-intensity Compton photon beam of LUXE can be used to search for physics beyond the Standard Model.}

\section{Strong-field QED}

The theory of Quantum Electrodynamics (QED) has been tested with unrivalled precision. Most experimental tests of QED to this day, however, consider QED only in the perturbative regime. LUXE~\cite{LUXECDR} aims to probe QED in a new regime of strong fields, created in interactions between the $16.5\,$GeV Eu.XFEL electron beam and a $40\,\textrm{TW}$ optical laser, as well as a secondary high-energy gamma photon beam and the laser. The Lorentz boost of the electrons in conjunction with the high-intensity laser create an electromagnetic field strength above the \textit{Schwinger limit}~\cite{Schwinger:1951} (in case of an electric field $E_{\textrm{cr}} =m^2_ec^3/(e\hbar)\approx 1.32\times10^{18}\,\textrm{V/m}$) \footnote{Here, $m_e$ denotes the electron mass, $c$ the speed of light in vacuum, $e$ the electron charge and $\hbar$ the reduced Planck constant.}, where QED becomes non-perturbative. Above the Schwinger limit, the work by the field over one Compton wavelength $\lambdabar=\hbar/(m_ec)$ of the probe electron is larger than the rest mass $m_e$. As a result, the QED vacuum becomes polarized and processes to all orders of the electron-laser coupling contribute to the scattering amplitude~\cite{Fedotov:2022ely}. \\

Two parameters govern the strong-field QED physics of LUXE: The \textit{classical nonlinearity parameter} $\xi=m_ec^2E/(\hbar\omega_LE_\textrm{cr})$, where $\omega_L$ is the laser wavelength, $E$ is the electromagnetic field strength, and $E_\textrm{cr}$ is the Schwinger critical field, quantifies the coupling between the laser background field and the probe particle. The probability of an n-th order electron-laser interaction process is therefore proportional to $\xi^{2n}$.  The \textit{quantum non-linearity parameter} $\chi=eE\lambdabar/(mc^2)$ characterizes the  field strength experienced by the probe particle in its rest frame in relation to the Schwinger critical field, as well as the recoil experienced by the probe particle emitting a photon. The two parameters are related via the equation $\chi=\xi\eta$, where $\eta=\gamma\hbar\omega_L(1+\cos\theta)/(m_ec^2)$ is the \textit{energy parameter}, $\gamma$ is the relativistic Lorentz factor, and $\theta$ is the electron-laser crossing angle ($\theta\approx17.2^\circ$ in LUXE).\\

\begin{figure}
\begin{minipage}{0.49\linewidth}
\centerline{\includegraphics[width=\linewidth]{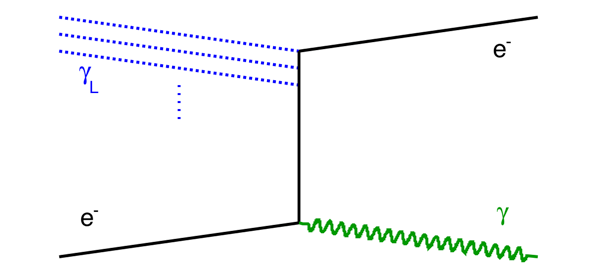}}
\end{minipage}
\hfill
\begin{minipage}{0.49\linewidth}
\centerline{\includegraphics[width=\linewidth]{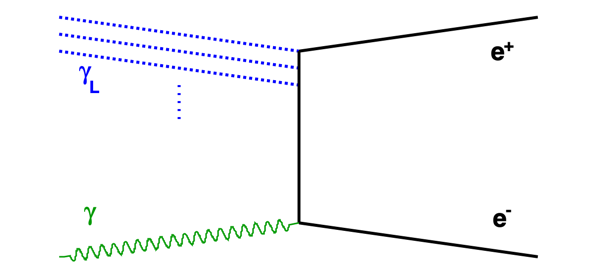}}
\end{minipage}
\hfill
\caption[]{Feynman diagrams of strong-field QED processes at LUXE, showing non-linear Compton scattering (left) and non-linear Breit-Wheeler pair production (right).}
\label{fig:feyn}
\end{figure}

\begin{figure}
\begin{minipage}{0.49\linewidth}
\centerline{\includegraphics[width=\linewidth]{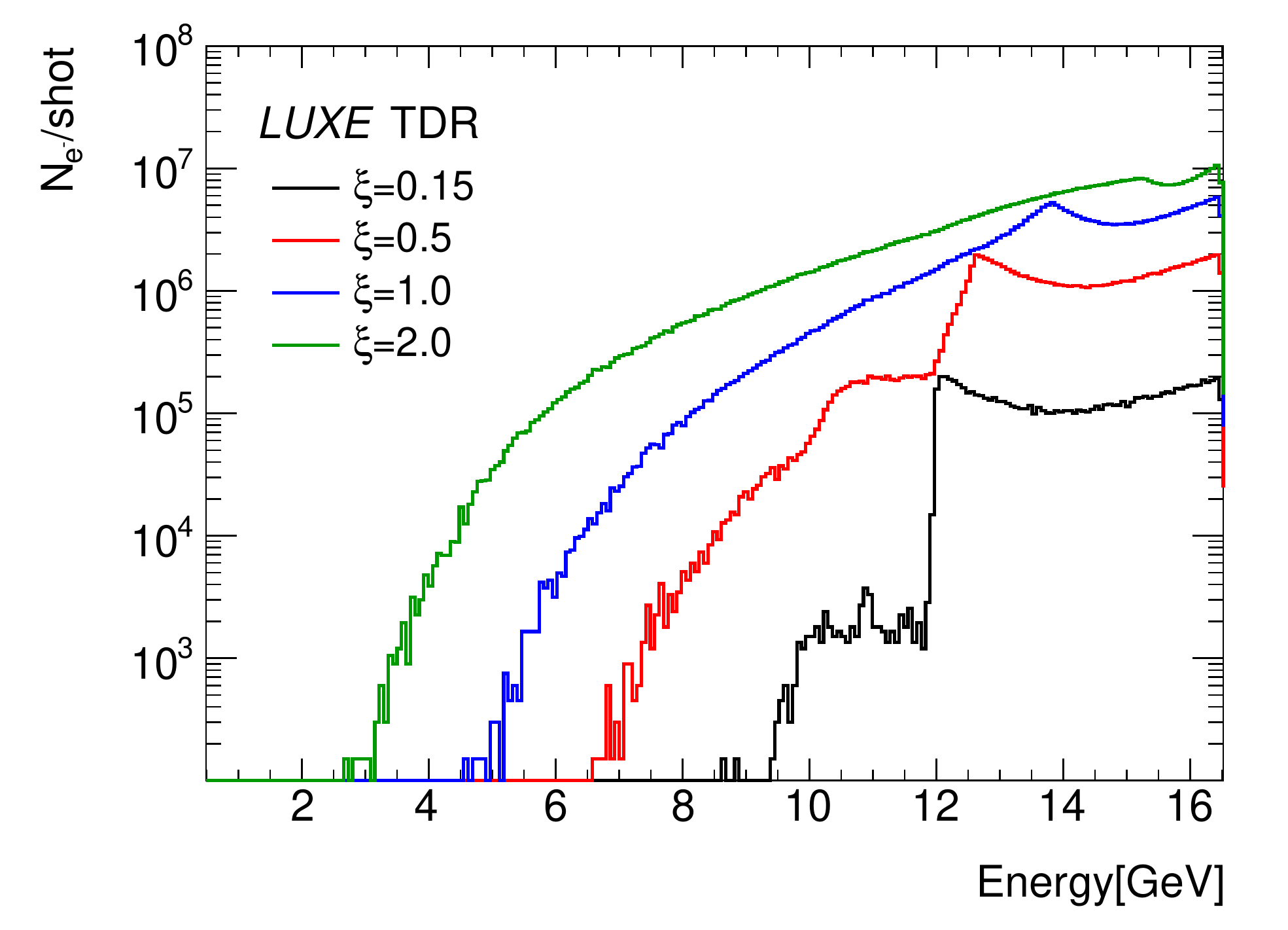}}
\end{minipage}
\hfill
\begin{minipage}{0.49\linewidth}
\centerline{\includegraphics[width=\linewidth]{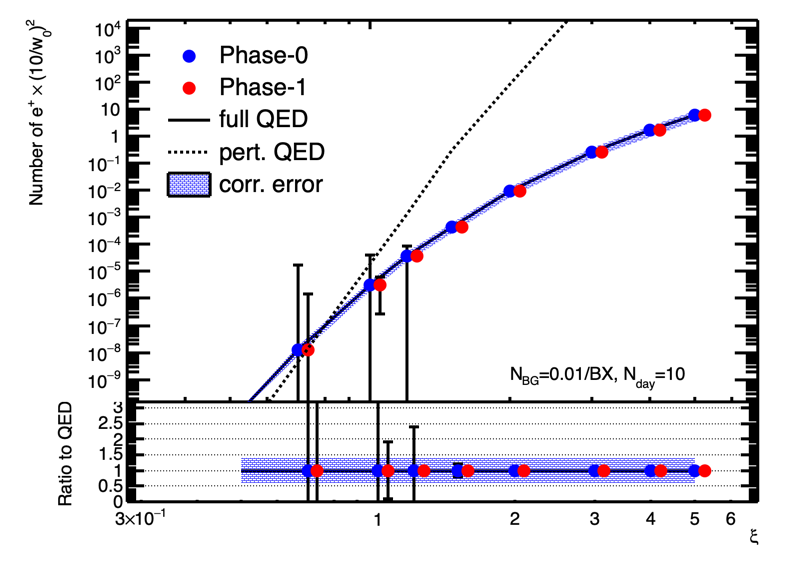}}
\end{minipage}
\hfill
\caption[]{Left: Simulated Compton electron energy spectrum for one electron-laser bunch crossing as a function of the laser intensity $\xi$ in the LUXE $e^-$-laser mode. Right: Breit-Wheeler positron rate as a function of $\xi$ in the LUXE $\gamma$-laser mode. Uncertainty corresponds to one week of data-taking.}
\label{fig:spectra}
\end{figure}

The main processes of strong-field QED probed by LUXE are \textit{non-linear Compton scattering} (see fig. \ref{fig:feyn} (left)) and \textit{non-linear Breit-Wheeler pair creation} (see fig. \ref{fig:feyn} (right)). In non-linear Compton scattering, the probe electron absorbs multiple laser photons and emits a single high-energy photon. The distinct feature of non-linear Compton scattering is the displacement of the Compton edge as function of the laser intensity parameter $\xi$, due to the fact that the probe electron acquires a larger effective mass $m_e^*=m_e\sqrt{1+\xi^2}$ in the laser field. Figure \ref{fig:spectra} (left) shows the Compton electron energy spectrum for different values of $\xi$ simulated with the \textsc{Ptarmigan} strong-field QED generator package~\cite{ptarmigan}.\\

Breit-Wheeler pair production is the field-induced creation of a physical electron-positron pair from the QED vacuum. In this process, a high-energy photon, produced in the LUXE $e^-$-laser collision mode via non-linear Compton scattering, or, in the LUXE $\gamma$-laser mode, stemming from a secondary high-energy photon beam, absorbs multiple laser photons and produces an electron-positron pair. This process has no classical equivalent, unlike the Compton process, therefore it is a direct probe of non-perturbative QED. The positron production rate, $\Gamma_\textrm{BW}$ is one of the main quantities of interest for LUXE. Figure \ref{fig:spectra} (right) shows the behaviour of $\Gamma_\textrm{BW}$  as a function of the laser intensity $\xi$. For $\xi\ll1$ the positron rate $\Gamma_\textrm{BW}$ follows a power law $\Gamma_\textrm{BW} \sim \xi^{2n}$. In the regime $\xi\gg 1$, a departure from the power-law characteristic occurs, where, for $\chi\ll 1$,  $\Gamma_\textrm{BW}$ scales according to:
\begin{equation}
    \Gamma_\textrm{BW}\propto\exp\left[-\frac{3}{8}\frac{1}{\omega_\gamma(1+\cos\theta)}\frac{E_\textrm{cr}}{E_\textrm{L}}\right].
    \label{eq:bwrate}
\end{equation}

In the fully non-perturbative regime, where $\xi\gg 1$ and $\chi\gg 1$, radiative quantum corrections to all orders contribute to $\Gamma_\textrm{BW}$. An equivalent to eq.~\ref{eq:bwrate} in the case where $\chi\gtrsim 1$ can be found in the following article~\cite{blackburn:king}.

\section{LUXE experimental setup}

\begin{figure}
\begin{minipage}{0.49\linewidth}
\centerline{\includegraphics[width=\linewidth]{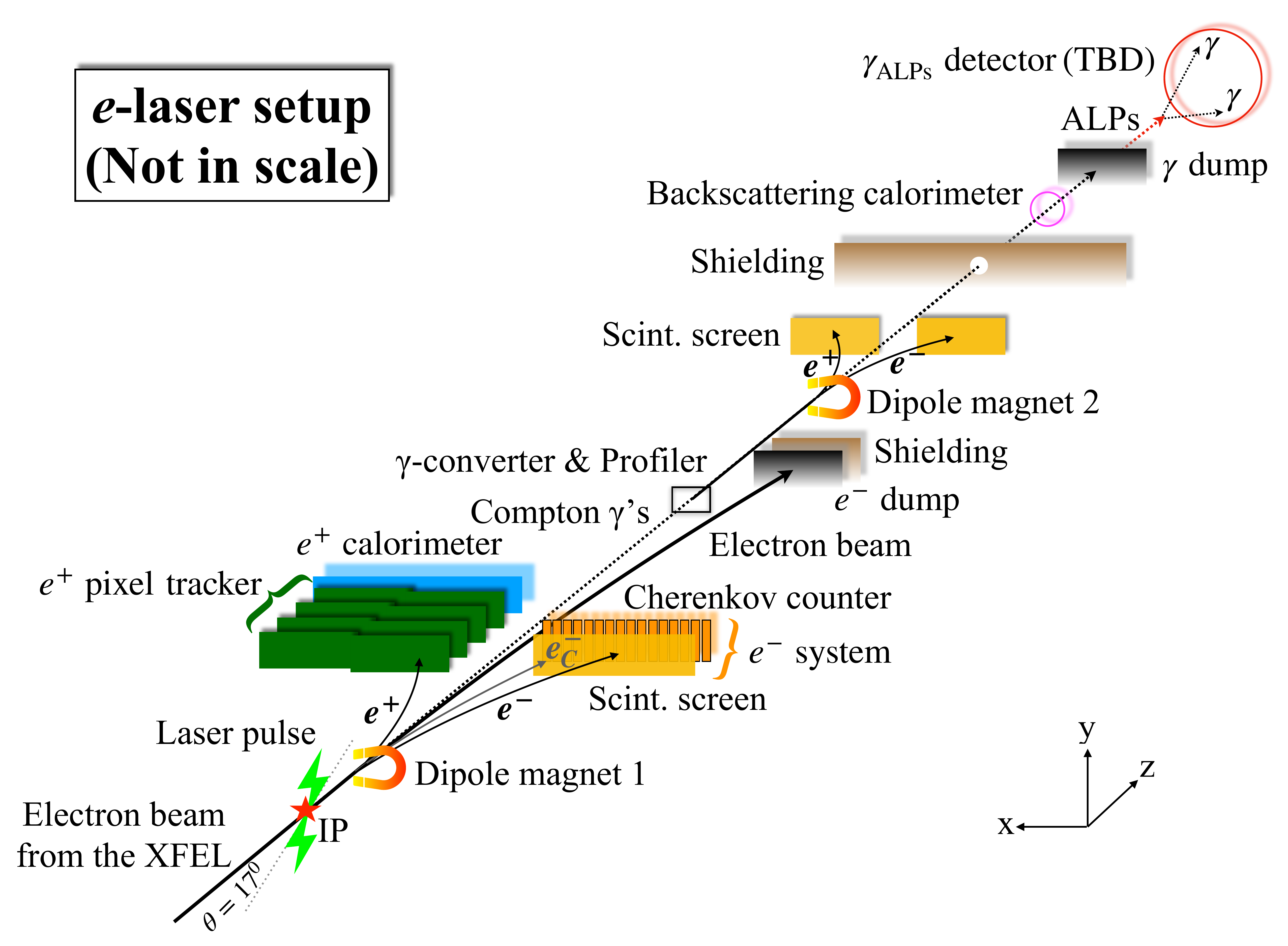}}
\end{minipage}
\hfill
\begin{minipage}{0.49\linewidth}
\centerline{\includegraphics[width=\linewidth]{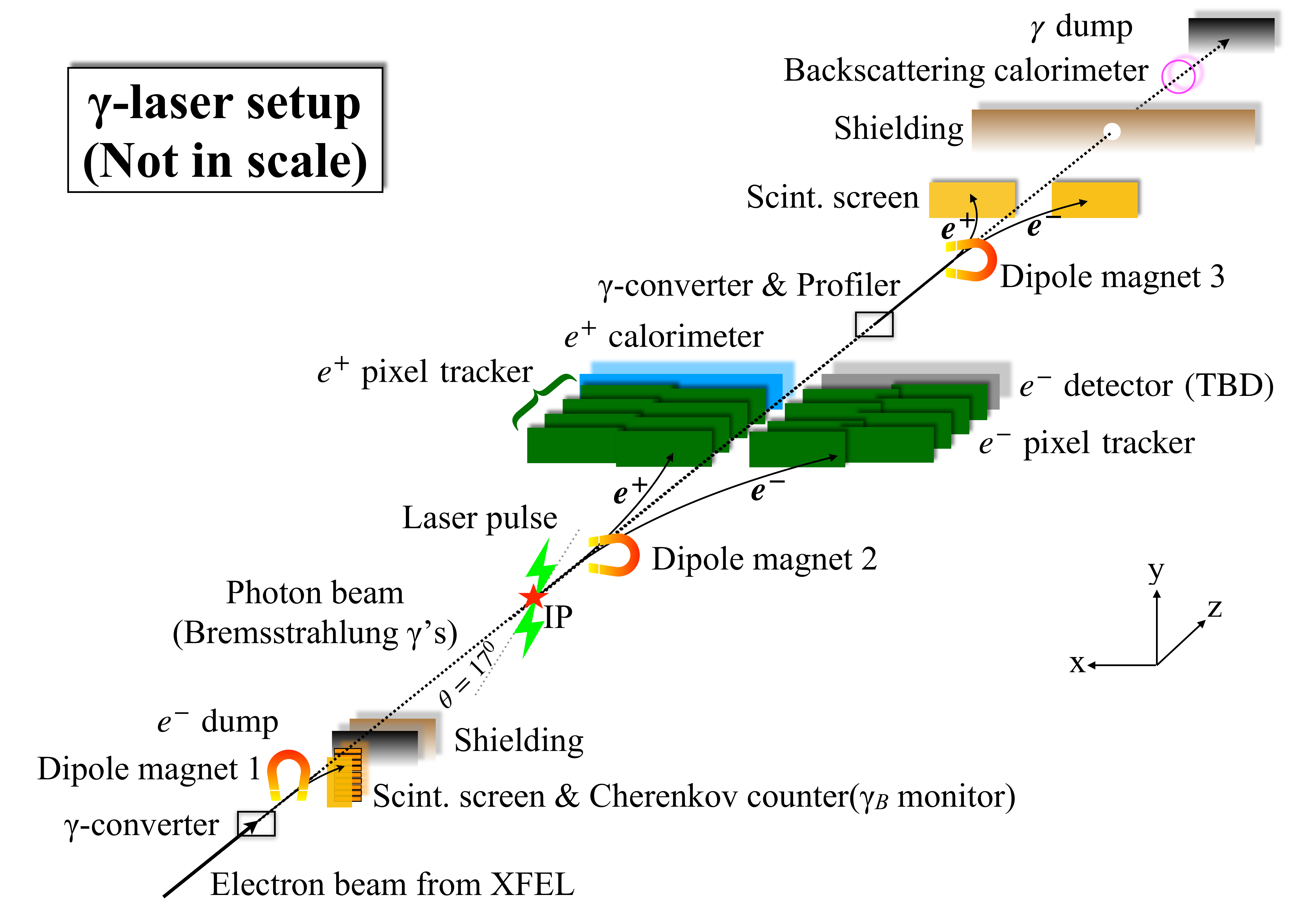}}
\end{minipage}
\hfill
\caption[]{Left: Schematic of the LUXE experimental setup in $e^-$-laser mode. Right: LUXE experimental setup in $\gamma$-laser mode.}
\label{fig:expsetup}
\end{figure}

Figure \ref{fig:expsetup} shows the LUXE experimental setup for the $e^-$-laser colision mode (left) and the $\gamma$-laser collision mode (right). In the $e^-$-laser mode, the $16.5\,\textrm{GeV}$ electron beam directly travels to the interaction point where it collides with the high-intensity laser. In the $\gamma$-laser mode, a secondary high-energy photon beam is created by impinging the electron beam on a Tungsten  Bremsstrahlung target, or, alternatively,by colliding the beam electrons with a low-intensity laser pulse to produce Compton photons with a narrow energy bandwidth.\\

The laser system foreseen for LUXE is a commercially sourced titanium-sapphire ($\lambda=800\,\textrm{nm}$) $40\,\textrm{TW}$ pulsed laser system (LUXE phase-0), which is upgradeable to $350\,\textrm{TW}$ (LUXE phase-1). The parameter space in terms of the nonlinearity parameters $\xi$ and $\chi$ that is accessible with the phase-0 (phase-1) laser setup is $\xi<7.9$ ($\xi<23.6$) and $\chi<1.5$ ($\chi<4.45$). One of the main challenges of LUXE is the required shot-to-shot stability of the laser ($<1\%$ variation, $<5\%$ intensity uncertainty), which is ensured by a dedicated suite of laser diagnostics.\\

The particle detectors in LUXE are used to detect electrons, positrons and photons and to measure their energy based on the transverse displacement in a dipole spectrometer. Since the particle rates per bunch-crossing (BX) in LUXE are vastly different, depending on the run mode, location in the experimental setup and laser intensity, dedicated technologies are used for each system. For the positron detection ($10^{-3}<N_{e^+}/BX<10^4$) a 4-layer silicon tracker in combination with a high-granularity electromagnetic calorimeter is used. For the Compton electron detection system the challenge is to reconstruct extremely high electron rates ($10^{3}<N_{e^-}/BX<10^8$). A finely segmented air-filled Cherenkov detector in combination with a scintillator screen read out by an optical camera system provides a robust solution. Finally, the photon energy spectrum is measured by a combination of three complementary detectors in the forward region of the experiment. Firstly, the gamma spectrometer partly reconverts the photon beam on a target into electron-positron pairs, which are subsequently analysed in a dipole spectrometer using a scintillator and camera system similar to the one used for the Compton-scattered electrons. Secondly, the gamma beam profiler studies the beam distribution in the transverse plane using sapphire strip detectors, in order to precisely determine the laser intensity $\xi$. Finally, a lead-glass calorimeter monitors backscattered particles from impinging the photons on the final dump, thereby measuring the total photon flux. In general, at least two complementary detector technologies are foreseen for each location in LUXE, to enable cross-calibration and the reduction of systematic uncertainties.\\

In addition to the LUXE strong-field QED program, the experimental setup can be extended to probe physics beyond the Standard Model (BSM) coupling to photons~\cite{LUXENPOD}. Rare BSM particles, such as axion-like particles (ALPs) could be produced in the interaction of the LUXE high-intensity Compton photon beam with the photon beam dump material, for example via Primakov production. A photon detector placed in a well-shielded area behind the beam dump can be used to search for ALPs decaying to two photons in the volume between the dump and the detector. The expected sensitivity of the LUXE extended setup is comparable to ongoing and planned experiments searching for ALPs. 

\section{Conclusion}

The aim of LUXE is to probe QED in a new regime of strong-fields, by studying collisions between the EU.XFEL electron beam, or a high-energy secondary gamma photon beam, with a high-intensity optical laser. Running LUXE as a collision experiment in continuous data-taking mode will enable precision measurements of strong-field QED processes, such as non-linear Compton scattering and Breit-Wheeler pair production. The laser system and particle detectors in LUXE are custom-designed in order to meet the physics goals. LUXE will likely be the first experiment to take precision measurements in a regime of QED never before explored in clean laboratory conditions and to study high-intensity laser collisions with real high-energy gamma photons.

\section*{Acknowledgments}

This work was in part funded by the Deutsche Forschungsgemeinschaft under Germany’s Excellence Strategy – EXC 2121 “Quantum Universe" – 390833306.\\

This work has benefited from computing services provided by the German National Analysis Facility (NAF).

\end{document}